\documentclass{article}%
\usepackage{a4wide}
\usepackage{amsmath}
\usepackage{amsfonts}
\usepackage{amssymb}
\usepackage{graphicx}%
\begin{document}

\title{Bose-Einstein transition temperature in a dilute repulsive gas}
\author{M.\ Holzmann$^{1}$, J.N.\ Fuchs$^{2}$, G.\ Baym$^{3}$, J.P.\ Blaizot$^{4}$,
and F. Lalo\"{e}$^{2}$\\$^{1}$ Laboratoire de Physique Th\'{e}orique des Liquides,\\UPMC, 4 place Jussieu, 75252 Paris, France\\$^{2}$ Laboratoire Kastler Brossel, \'{E}cole Normale Sup\'{e}rieure,\\24 rue Lhomond, 75005 Paris, France\\$^{3}$ University of Illinois at Urbana-Champaign, \\1110 W. Green St., Urbana, IL 61801, USA\\$^{4}$ CEA-Saclay, Service de Physique Th\'{e}orique, 91191 Gif-sur-Yvette,
France }
\maketitle

\begin{abstract}
We discuss certain specific features of the calculation of the critical
temperature of a dilute repulsive Bose gas. Interactions modify the critical
temperature in two different ways.\ First, for gases in traps, temperature
shifts are introduced by a change of the density profile, arising itself from
a modification of the equation of state of the gas (reduced compressibility);
these shifts can be calculated simply within mean field theory.\ Second, even
in the absence of a trapping potential (homogeneous gas in a box), temperature
shifts are introduced by the interactions; they arise from the correlations
introduced in the gas, and thus lie inherently beyond mean field theory - in
fact, their evaluation requires more elaborate, non-perturbative,
calculations. One illustration of this non-perturbative character is provided
by the solution of self-consistent equations, which relate together
non-linearly the various energy shifts of the single particle levels $k$.
These equations predict that repulsive interactions shift the critical
temperature (at constant density) by an amount which is positive, and simply
proportional to the scattering length $a$; nevertheless, the numerical
coefficient is difficult to compute.\ Physically, the increase of the
temperature can be interpreted in terms of the reduced density fluctuations
introduced by the repulsive interactions, which facilitate the propagation of
large exchange cycles across the sample.

\end{abstract}

\begin{center}
{\Large Introduction}
\end{center}

The calculation of the effects of interactions in dilute gases is often
considered as a classical textbook problem, which is generally solved with the
help of cluster techniques in statistical mechanics; well known examples are
for instance the second virial correction to the pressure, the heat capacity,
or to the magnetic susceptibility for a gas of spin $1/2$ particles
\cite{Huang, Pathria}.\ For gases at low temperatures, all these corrections
are expressed in terms of a single parameter, the scattering length $a$, so
that the result is simple. At first sight, the calculation of the first
correction to the critical Bose-Einstein condensation (BEC)\ temperature in a
dilute repulsive Bose gas seems to raise a similar problem.\ Nevertheless, its
solution was not well understood until recently, as illustrated by a large
collection of contradictory results in the literature (see \cite{BBHLV-2} and
references contained).\ Several reasons underlie the confusion.\ First, for a
homogeneous gas at low temperatures, it turns out that a mean field treatment
of the interactions leads to an exactly zero shift of the critical temperature
(the critical value of the chemical potential is shifted, but this does not
affect the critical temperature); combining mean field theory with additional
- and not necessarily well controlled - approximations can then lead to
results which are approximation dependent, providing arbitrary $a$ power
variations and even sign.\ The second reason is that the problem is actually
not as simple as it looks.\ It is essentially non-perturbative at several
levels \cite{BBHLV-2, BBHLV-1}, even if the leading correction term which
emerges from the calculations ends up being simply linear in $a$; we discuss
below in more detail this peculiar feature of the calculations.\ In
retrospect, this difficulty is actually not so surprising, since the
properties of a Bose gas are non-analytic around $a=0$, where the system is at
the border of stability (a Bose gas collapses at condensation as soon as $a$
becomes negative).\ The linearity in $a$ is therefore far from obvious, and
indeed it turns out that the next correction \cite{BBHL} is not simply $\sim
a^{2}$ but $\sim a^{2}\log a$, clearly manifesting the non-analytic character
of the problem.

The purpose of the present article is to return to the considerations of
\cite{BBHLV-2} and \cite{BBHLV-1}, with more detailed discussion of some of
their physical aspects. In particular, we wish to emphasize that the linearity
in $a$ of the leading term in the correction to the critical temperature may
be missed if a non self-consistent theory is used to calculate the effect;
indeed, some calculations found in the recent literature and based on non
self-consistent models provide different results, first corrections $\sim
a\log a$ for instance. But, before we come to this point, in order to avoid
any confusion, we carefully distinguish between two effects which are
discussed in this context; the former takes place in traps only, the latter
also in a uniform gas contained in a box.

\section{Compressibility effects in a trap (mean field)\label{compressibility}%
}

For a Bose gas of $N$ atoms contained in a trap at a given temperature,
Bose-Einstein condensation first sets in at the point where the density is
maximum.\ Repulsive interactions tend to make the density more uniform, and
hence will in general lower the density at this point, so that they will
decrease the transition temperature with respect to an ideal gas.\ Several
authors have studied this effect; for a review, see ref.\ \cite{DGPS}
(\S \ V-B) and references therein.\ \ Here, for the sake of simplicity, we
limit ourselves to the discussion of isotropic traps and to the thermodynamic
limit\footnote{The thermodynamic limit in a trap is defined as the limit where
the number $N$ of particles tends to infinity, the frequency $\omega$ of the
trap goes to zero, and the product $N\omega^{3}$ remains constant.}.\ This
excludes finite size effects, which are also discussed in \cite{DGPS}; a
finite size system does not undergo a sharp phase transition and thus the
critical temperature, and its shift, depend on a somewhat arbitrary definition
of the critical temperature\footnote{In an ideal gas, the number of particles
in excited states $N_{e}$ can not exceed a maximum value $N_{e}^{\max}$,
obtained when the chemical potential is equal to the single particle ground
state energy.\ The usual definition of the critical value $N_{c}$ for the
total number of particles $N$ in a finite trap is $N_{c}=N_{e}^{\max}$. This
convention is convenient, but remains somewhat arbitrary, since it relates the
value of one physical quantity, $N$, to the upper limit of another physical
quantity $N_{e}$.\ It therefore conveys the idea of $N_{e}$ increasing until
it reaches its absolute upper limit, and then saturating.\ In reality, a large
number of particles has already accumulated in the ground state when
$N=N_{e}^{\max}$, and $N_{e}$ actually never reaches this upper limit (except
in the limit $N\rightarrow\infty$).
\par
Another definition of the critical point may be obtained by plotting the
number $N_{0}$ of particles in the lowest state as a function of temperature
(at constant $N$), and taking the inflexion point to define the critical
temperature.\ This is in a sense more physical, but still arbitrary, and this
different convention leads to different values of the finite size effects.}.
In the thermodynamic limit, this arbitrary character disappears; the width of
the trap becomes large so that the effects of the trapping potential may be
treated within a semi-classical approximation.\ Around the center of the trap,
one can then completely ignore the effects of the potential, and the
condensation is obtained when, locally, the critical conditions are
reached.\ This happens when the local degeneracy parameter, the product
$n(0)\lambda^{3}$, is equal to the critical value of this parameter in a
homogeneous system, where $n(0)$ is the number density at the center of the
trap and $\lambda=h/\sqrt{2\pi mk_{B}T}$ the thermal wavelength.\ For an ideal
gas, this value is $\zeta_{3/2}\simeq2.61..$; for an non ideal gas, it is
slightly shifted, see \S \ \ref{correlation}. The question now is to relate
$n(0)$ to $N$ in order to obtain its critical value as a function of the
temperature, or conversely the critical temperature as a function of $N$.

\subsection{Equation of state}

In a gas in equilibrium, the local density of the gas adjusts locally so that
the pressure gradient in the gas compensates exactly for the force exerted by
the external potential. The equation of state of the gas therefore plays an
essential role in the determination of the density profile in a trap. A good
approximation is provided by mean field theory, where the local density of the
gas is written as:%
\begin{equation}
n=\frac{1}{\lambda^{3}}~g_{3/2}\left[  z=\exp\beta(\mu-\Delta\mu)\right]
+n_{0} \label{1}%
\end{equation}
with:\
\begin{equation}
\beta\Delta\mu=4a\lambda^{2}n \label{2}%
\end{equation}
where $\beta=1/k_{B}T$ is the inverse temperature, $g_{3/2}(z)$ the usual Bose
function \cite{Pathria} and $n_{0}$ the density of condensed
particles.\ Figure 1-a shows the variations of the density\footnote{We take
this opportunity to note that fig.\ 2 of \cite{BBHLV-2} is inaccurate: at the
critical value $\mu_{c}$ of the chemical potential, the curve should not have
a kink, but a continuous slope.} as a function of the chemical potential $\mu
$, assuming that the gas is repulsive ($a>0$) - see for instance figure 10 of
ref.\ \cite{HGL} for a geometrical construction of this curve.\ Integration
over the chemical potential provides the pressure (fig.\ 1-b); eliminating
$\mu$ leads to the equation of state in mean field approximation (fig. 1-c).
For comparison, the curves are also shown for the ideal gas ($a=0$).\ As
expected, repulsive interactions tend to increase the pressure of the gas and
to reduce its compressibility. \begin{figure}[ptb]
\begin{center}
\includegraphics[
height=4.5cm,
width=13cm
]{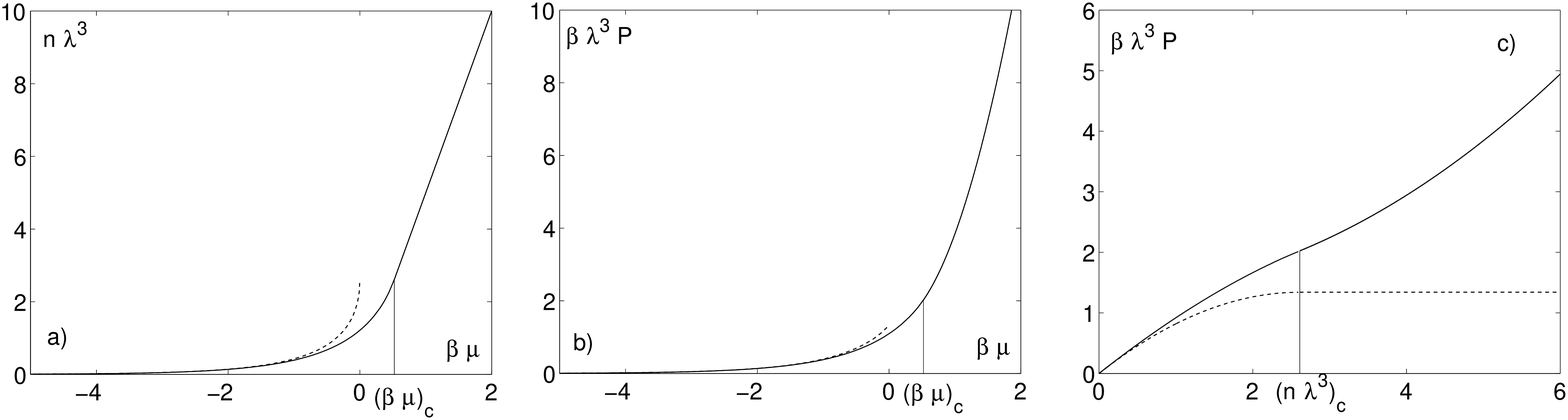}
\end{center}
\caption{Figure (a) shows the number density $n$ in the gas as a function of
the chemical potential $\mu$, at constant temperature $T$ ($n$, measured in
unities of $\lambda^{-3}$, is dimensionless). This curve is obtained within
simple mean field theory; the critical value of $\mu$ is $\mu_{c}$; beyond
this value, the slope of the curve is inversely proportional to the scattering
length $a$. Figure (b) shows the pressure $P$ in the gas, which is the
integral of $n$ over $\mu$ .\ Figure (c) shows the equation of state relating
$P$ to $n$, after elimination of $\mu$; the critical value of the density is
$n_{c}$; at this point, the compressibility of the gas is proportional to
$1/a$. Dashed lines refer to the ideal gas ($a=0$).}%
\label{fig-1}%
\end{figure}

\subsection{Total number of particles in a trap vs. density at the
center\label{total}}

In a trap that is sufficiently large (thermodynamic limit), the effects of the
external potential can be treated semi-classically, which allows one to extend
equations (\ref{1},\ref{2}) to inhomogeneous systems.\ We now assume that the
gas is not Bose condensed and define the effective chemical potential
$\mu_{\text{eff}}$ by:%
\begin{equation}
\mu_{\text{eff}}(\mathbf{r})=\mu-V(\mathbf{r})-4a\lambda^{2}~n(\mathbf{r}%
)/\beta\label{3-2}%
\end{equation}
and find:%
\begin{equation}
n(\mathbf{r})\lambda^{3}=~g_{3/2}\left[  z=e^{\beta\mu_{\text{eff}}%
(\mathbf{r})}\right]  \label{3-1}%
\end{equation}
But equation (\ref{3-1}) can be inverted as:%
\begin{equation}
\mu_{\text{eff}}=\beta^{-1}J(n\lambda^{3}) \label{3-3}%
\end{equation}
where $J$ is the logarithm of the inverse function of $g_{3/2}$. By
differentiation, one obtains:%
\begin{equation}
\bigtriangledown\mu_{\text{eff}}=\beta^{-1}\lambda^{3}~J^{\prime}(n\lambda
^{3})~\bigtriangledown n \label{3-4}%
\end{equation}
where $J^{^{\prime}}$ is the derivative of $J$. Combined with (\ref{3-2}),
this equation yields:%
\begin{equation}
-\beta\bigtriangledown V(\mathbf{r})=\left[  \lambda^{3}~J^{\prime}%
(n\lambda^{3})+4a\lambda^{2}\right]  ~\bigtriangledown n(\mathbf{r}) \label{4}%
\end{equation}
which relates the local variations of the potential and of the density.\ In
general, the term in $4a\lambda^{2}$ on the right side of this equation is a
small correction to the term in $J^{\prime}$, so that the relation between the
two gradients depends only weakly on $a$.\ Nevertheless, in regions of space
where the gas is almost condensed, the derivative $J^{^{\prime}}$ becomes very
small and, for a given $\bigtriangledown V$, the density gradient
$\bigtriangledown n$ becomes much larger and strongly $a$ dependent; this
phenomenon can take place at the center of the trap, as we discuss in more
detail in \S \ \ref{combined}.

Equation (\ref{4}) allows one to calculate $n(\mathbf{r})$ at every point of
the gas as a function of $n(0)$ and, through an $\mathbf{r}$ integration, of
the total number of particles $N$.\ At the critical point, $n(0)$ is fixed by
the condition $n(0)\lambda^{3}$ $=\zeta_{3/2}$ \ in the mean field
approximation\footnote{In all this article, we assume that the interactions
can be described by a single parameter $a$, and therefore have no $\mathbf{k}$
dependence.\ In this case, one can easily show \cite{BBHLV-1}\cite{HGL} that,
within mean field approximation, the critical value of the degeneracy
parameter is unaffected by the interactions.}, so that this calculation
provides a relation between the critical values of $N$ and $T$. For a harmonic
trap, where the size of the single particle ground state is $a_{ho}%
=\sqrt{\hbar/m\omega}$, Giorgini et al. \cite{GPS} show that, to lowest order,
this critical temperature is shifted with respect to its ideal gas value
$T_{c}^{0}$ by an amount $\delta T_{c}$ given by:%
\begin{equation}
\frac{\delta T_{c}}{T_{c}^{0}}=-1.3~\frac{a}{a_{ho}}N^{1/6} \label{5}%
\end{equation}
Since $N^{1/6}/a_{ho}\sim(N\omega^{3})^{1/6}$, the relative change of
temperature remains finite in the thermodynamic limit.\ Significant critical
temperature shifts can indeed be observed experimentally; for instance, recent
experiments by Gerbier et al. \cite{Gerbier} have reported observations of
relative shifts as large as 10\% in a Rb gas, in good agreement with
(\ref{5}); early Monte Carlo calculations on a trapped Bose gas also predicted
a temperature shift dominated by the mean field \cite{K} \cite{HKN}.

A few remarks may be useful at this stage:

(i) the shift $\delta T_{c}$ describes the case where the total number of
atoms $N$ is kept constant; if, instead, the density of the gas at the center
of the trap were kept constant, the shift would vanish (within mean field
theory).\ Within a factor, $N$ and $n(0)$ are conjugate
variables\footnote{When the potential is treated semi-classically, one can
show that \ $N=g_{3/2}^{\prime}(z)\times n(0)~\partial LogZ/\partial n(0)$
where $Z$ is the partition function with the usual notation $z=\exp\beta\mu$%
.}, the former extensive, the latter intensive, somewhat similar to energy and
temperature in thermodynamics.\ Neither of them is measured directly in
experiments, since optical measurements provide the \textquotedblleft column
density\textquotedblright\ (density integrated along the line of sight), an
intermediate physical quantity.\ Conceptually, there is nevertheless a
preference for using $n(0)$ as the relevant variable: while the calculation of
the shift $\delta T_{c}$ at constant $N$ $\ $gives different results for all
possible forms of traps (parabolic, quartic, or even more complicated as that
shown schematically in figure 2), in the thermodynamic limit, all lead to the
same value of $n(0)$; this result is clearly more universal.

(ii) the physics involved in $\delta T_{c}$ is only remotely related to the
physics of the Bose-Einstein transition; it is more closely associated with
the effects of interactions on the density profile in trap, which is in turn
determined by the equation of state. In other words, the calculation of $N$
involves many atoms which play actually no role in the condensation, because
they are in regions of space where the density is too low, almost or
completely in a classical regime. This is particularly obvious for the trap of
figure 2\thinspace\ where, clearly, all the atoms in the external part of the
trap are counted in $N$, with a weight proportional to the volume of this
part, while they play no role whatsoever in the condensation phenomenon.

\begin{figure}[ptb]
\begin{center}
\includegraphics[
height=1.2237in,
width=2.1326in
]{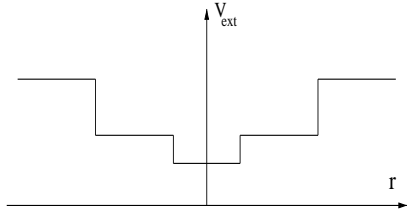}
\end{center}
\caption{Example of a shape of a trap where only the central part plays a role
in Bose-Einstein condensation; all the atoms in the other part are counted in
the total number of particles $N$ but do not take part in the phenomenon.}%
\label{fig-2}%
\end{figure}

\section{Correlation effects in a uniform gas}

\label{correlation}

We now come to the central part of this article and discuss another shift of
the critical temperature of totally different physical origin.\ This effect is
not related to an external trapping potential but emerges from correlations
between the particles; it is actually most conveniently calculated for a
uniform gas contained in a box. As recalled in the introduction, the study of
this shift has a long history of contradictory results, predicting either an
increase or a decrease of the critical temperature, as well as variations with
different powers of $a$ (see \cite{BBHLV-2}). It is now understood that
correlation effects taking place in the non condensed gas just above
transition lead to an increase of the critical temperature $T_{c}$, which is
proportional to $a$ in leading order:%
\begin{equation}
\frac{\Delta T_{c}}{T_{c}^{0}}\simeq~c~an^{1/3} \label{5-2}%
\end{equation}
where $c$ is a numerical positive coefficient. In \cite{BBHLV-2, BBHLV-1},
linearity emerges as an exact result from a scaling analysis of the complete
perturbation series.

In this text, we give another version of the argument leading to this
linearity, with more emphasis on simple properties of non-linear
self-consistent equations, and on the physical mechanism they contain.\ Behind
the increase of $T_{c}$ is a modification of the effective energies that
determine the populations of the particles with low momenta. Within the
framework of the non-linear self-consistent equations introduced in \cite{HGL,
BBHLV-2, BBHLV-1}, and without elaborate calculations, we summarize the basic
ideas that lead to the shift of the transition temperature, and comment on
various points involved.

\subsection{Non-linear equation for the single particle
energy\label{non-linear}}

In a non condensed gas, the diagonal elements $\rho_{k}$ of the single
particle density matrix can be written as:%
\begin{equation}
\rho_{\mathbf{k}}=~\left[  e^{\beta(e_{\mathbf{k}}-\mu+\Sigma_{\mathbf{k}}%
)}-1\right]  ^{-1} \label{6}%
\end{equation}
where $\beta$ is the inverse temperature, $\mu$ the chemical potential and the
effective energy is the sum of the free particle kinetic energy $e_{\mathbf{k}%
}=\hbar^{2}\mathbf{k}^{2}/2m$ and the energy shift $\Sigma_{k}$ introduced by
the interactions. As in \cite{HGL}, we write\footnote{In Ursell theory,
equations (\ref{6}) and (\ref{7}) are naturally obtained in a self consistent
second order approximation.\ Note that the energy shifts $\Sigma_{\mathbf{k}}$
are then distinct from the self energies defined in Green's function theory;
in particular, they provide no information on the evolution of the
quasiparticles, but just the static populations $\rho_{\mathbf{k}}$. See ref.
\cite{FHL} for a discussion of the appearance of exponentials exp$\beta
\Sigma_{\mathbf{k}}$ in Ursell operator theory.
\par
In Green's function formalism, equations (\ref{6}) and (\ref{7}) can also be
obtained with the help of additional approximations.\ For instance, one can
neglect of the Matsubara frequency dependence of the self energies, which
allows one to reconstruct in $\rho_{\mathbf{k}}$ the Bose-Einstein
distribution function appearing in (\ref{6}), and recover the same
equations.}:
\begin{equation}
\beta\Sigma_{\mathbf{k}}=4a\lambda^{2}n-8\left(  \frac{a}{\lambda}\right)
^{2}\left(  \frac{\lambda}{2\pi}\right)  ^{6}\int d^{3}k^{^{\prime}}%
\rho_{\mathbf{k}^{^{\prime}}}\int d^{3}q~\rho_{\mathbf{k}+\mathbf{q}}%
~\rho_{\mathbf{k}^{^{\prime}}-\mathbf{q}} \label{7}%
\end{equation}
This equation is not exact, but a self-consistent \textquotedblleft one bubble
approximation\textquotedblright\ to $\Sigma_{\mathbf{k}}$ ; it nevertheless
allows a qualitative discussion of the properties of the more general theory
and provides a reasonable estimate of the leading correction for $\Delta
T_{c}$ in the homogeneous system. The first term on the right side of
(\ref{7}) is simply the mean field term; in the low temperature regime where
the interactions are described in terms of the scattering length $a$ only, it
depends only on the density:
\begin{equation}
n=\int\frac{d^{3}k}{(2\pi)^{3}}~\rho_{\mathbf{k}} \label{8}%
\end{equation}
Because this mean field is independent of $\mathbf{k}$, it has no effect
whatsoever on the critical density (at a given temperature). The second term
is more interesting; it corresponds to the effect of correlations due to
interactions in the gas, and introduces a $\mathbf{k}$ dependence of the
energy shift $\Sigma_{\mathbf{k}}$. We will see that this shift tends to
\textquotedblleft harden\textquotedblright\ the free particle spectrum around
$\mathbf{k}=0.$

The condition for Bose-Einstein condensation reads:
\begin{equation}
\mu=\Sigma_{0} \label{8-2}%
\end{equation}
so that, in terms of the \textquotedblleft spectrum\textquotedblright\ at the
critical point:%
\begin{equation}
W(\mathbf{k})=\beta\left[  e_{\mathbf{k}}+\Sigma_{\mathbf{k}}-\Sigma
_{0}\right]  \label{w}%
\end{equation}
eqs (\ref{6}) and (\ref{7}) provide the following self-consistent equation:
\begin{equation}%
\begin{array}
[c]{ll}%
W(\mathbf{k}) & \displaystyle=\beta e_{\mathbf{k}}-8\left(  \frac{a}{\lambda
}\right)  ^{2}\left(  \frac{\lambda}{2\pi}\right)  ^{6}\int d^{3}k^{^{\prime}%
}\int d^{3}q\\
& \displaystyle\frac{1}{e^{W(\mathbf{k}^{^{\prime}})}-1}~\frac{1}%
{e^{W(\mathbf{k}^{^{\prime}}-\mathbf{q})}-1}\left[  \frac{1}{e^{W(\mathbf{k}%
+\mathbf{q})}-1}-\frac{1}{e^{W(\mathbf{q})}-1}\right]
\end{array}
\label{8-3}%
\end{equation}
From the spectrum we can obtain the populations $\rho_{\mathbf{k}}$ and the
critical density $n_{c}$.\ Actually, what we wish to obtain is the change
$\Delta n_{c}$ of the critical density (at constant temperature) with respect
to the ideal gas value $n_{c}^{0}=\zeta_{3/2}\lambda^{-3}$.\ We assume that
only small values of $\mathbf{k}$ contribute to this change (we discuss why
this is true below); since $W(\mathbf{k}=\mathbf{0})$ vanishes, we can replace
$\left[  \exp W(\mathbf{k})-1\right]  ^{-1}$ by $1/W(\mathbf{k})$ to obtain:
\begin{equation}
\Delta n_{c}\simeq\int\frac{d^{3}k}{(2\pi)^{3}}\left[  \frac{\beta
e_{\mathbf{k}}-W(\mathbf{k})}{\beta e_{\mathbf{k}}~W(\mathbf{k})}\right]
\label{nc}%
\end{equation}
to leading order. Note that, as mentioned above, the mean-field term does not
enter the problem anymore, and therefore does not influence the critical
density or temperature.

At this point, it is easy to convert the change of density (\ref{nc}) at
constant temperature into a change of temperature at constant density. The
critical condition in the presence of interactions can be written as:%
\begin{equation}
n\lambda^{3}=\frac{nh^{3}}{\left(  2\pi mk_{B}T\right)  ^{3/2}}=\zeta
_{3/2}+\lambda^{3}\Delta n_{c}(a,T) \label{g}%
\end{equation}
where $\Delta n_{c}(a,T)$ is given by (\ref{nc}).\ Equation (\ref{g}) defines
the new critical line in the temperature-density plane (fig. 3).\ Any point on
the initial critical line (ideal gas) can be moved by a first order change of
either the density, or the temperature, and reach a point on the new critical
line; the only condition is that the change of the product $n\lambda^{3}$
should be equal to $\lambda^{3}\Delta n_{c}(a,T)$ to first order\footnote{To
first order, we can ignore the difference between $\Delta n_{c}(a,T+\Delta T)$
and $\Delta n_{c}(a,T)$.}. But the same change of $n\lambda^{3}$ can be
obtained, either when $\Delta n=0$ with a change of temperature $\Delta T\,$,
or when $\Delta T=0$ with a change of density $\Delta n$.\ In fact, for any
motion in the temperature-density plane, the relative variation of the left
side of (\ref{g}) is given by:%
\begin{equation}
\frac{\Delta(n\lambda^{3})}{(n\lambda^{3})}=\frac{\Delta n}{n}-\frac{3}%
{2}\frac{\Delta T}{T} \label{tc}%
\end{equation}
The same relative variation can therefore be obtained, either with a relative
change $\Delta n/n$ at constant $T$, or with a relative change $\Delta
T/T$\ at constant density, provided the ratio between the two changes is
$-2/3$. Finally, the leading term of the change of critical temperature
$\Delta T_{c}$ is given by:%
\begin{equation}
\frac{\Delta T_{c}}{T_{c}^{0}}\simeq-\frac{2}{3n_{c}^{0}}\int\frac{d^{3}%
k}{(2\pi)^{3}}\left[  \frac{\beta e_{\mathbf{k}}-W(\mathbf{k})}{\beta
e_{\mathbf{k}}~W(\mathbf{k})}\right]  \label{tc-2}%
\end{equation}

\begin{figure}[ptb]
\begin{center}
\includegraphics[
height=3cm,
]{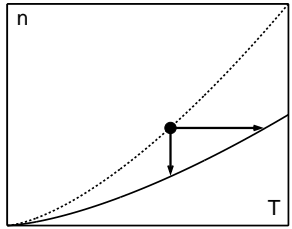}
\end{center}
\caption{Critical line in the density-temperature plane, for the ideal gas
(upper line) and for the interacting gas (lower line). Starting from a point
on the first critical line, arrows indicate how a change of density at
constant temperature, or conversely, move the point to the new line, as
discussed in the text.}%
\end{figure}\qquad\qquad

We will see in \S \ \ref{scl} that the change of the spectrum at the critical
point manifests itself primarily in a small $k$ region centered at the origin
and of width $\Delta k$ given by:%
\begin{equation}
\Delta k=k_{c}\sim\frac{a}{\lambda^{2}} \label{9}%
\end{equation}
In this region, the correction $\Sigma_{\mathbf{k}}-\Sigma_{0}$ is comparable
to the free particle energy $e_{\mathbf{k}}$, or may even dominate in
$W(\mathbf{k})$; this property is also discussed in detail in \cite{BBHLV-2,
BBHLV-1}.\ The physical origin of the change of energy is simple: the atoms
with very low velocities are extremely sensitive to even very small effects of
the interaction potential.\ They can therefore rearrange themselves in space
to minimize their repulsive energy, reaching a smaller value than the mean
field prediction.\ On the other hand, the atoms with higher $k$ values than
$\Delta k$ have too much kinetic energy to do so.

This rearranging process can be regarded as an analogue, in terms of
interaction potential and correlations, of the effect of a weak external
potential on particle positions discussed by Lamb et al. \cite{Lamb}. These
authors point out that the character of the Bose-Einstein condensation is
qualitatively modified even by a weak external potential, transferring it from
momentum space to real space.\ Here we have another illustration of the
extreme sensitivity of a gas just above transition to any potential, but in
the space of relative positions instead of ordinary position of the particles.

\subsubsection{Non self-consistent approximation: spurious
logarithms\label{spurious}}

The non-linear equations (\ref{6},\ref{7}), or (\ref{8-3}), are not easy to
solve.\ One can start by assuming that $\Sigma_{\mathbf{k}}$ provides only a
small correction to the free particle spectrum, so that the right side of
(\ref{7}) can be calculated with $\Sigma_{\mathbf{k}}=0$, which provides a
first approximation for $\Sigma_{\mathbf{k}}$ - this operation can be regarded
as a first iteration of an infinite process leading to a fully self-consistent
solution. The critical condition (\ref{8-2}) then gives an implicit equation
in $\mu$ to determine the critical value of the chemical potential (still
within this first iteration). For this value of $\mu$, one can then use
(\ref{6}) to calculate the new populations $\rho_{\mathbf{k}}$, and finally
$n$ by integration over $k$. This is precisely what was done in \cite{HGL} for
values of $a/\lambda$ ranging from $10^{-3}$ to $10^{-2}$, leading to the
following value of the coefficient $c$ (obtained by dividing the relative
change of $T_{c}$ by $an^{1/3}$):
\begin{equation}
c\simeq0.7 \label{10}%
\end{equation}

One can of course also make two (or more) iterations: again choose $\mu$,
calculate the first approximation for $\Sigma_{\mathbf{k}}$, the first
approximation for $\rho_{\mathbf{k}}$, inject them into ~(\ref{7}) to obtain
the second approximation for $\Sigma_{\mathbf{k}}$, and then only determine
$\mu$ by the condensation condition.\ The second approximation for
$\rho_{\mathbf{k}}$ eventually provides the critical density by integration
over $k$. This procedure leads to a progressive hardening of the spectrum, as
illustrated by Fig.~15 of \cite{HGL}, and typically to higher values of $c$
than suggested by the first iteration approximation, eq.~(\ref{10}).

But what is the validity of these calculations?\ Knowing that the presently
accepted value is $c\simeq1.3$ \cite{AM,KPS}, at first sight one could
consider this calculation a success, especially in view of its
simplicity.\ However, success may be pure coincidence!\ In fact, an analytical
study of the single iteration procedure shows that it does not lead to a
linear dependence of $\Delta T_{c}$ in $a$, but actually contains spurious
logarithms in $a$; these logarithms will dominate in the limit $a\rightarrow0$
and destroy the linearity - we come back to this point in \S \ \ref{IR}, but
see also \S \ 5.1 of \cite{BBHLV-2}. In other words, if the numerical
calculations leading to (\ref{10}) had been made with a different range for
the parameter $a/\lambda$, a different value for the coefficient $c$ would
have been obtained, as illustrated by the table of \S ~5.1 in \cite{BBHLV-2}.
The validity of the first iteration leading to eq.\ (\ref{10}) is therefore
difficult to assess a priori; see \cite{Stoof-2} for another example of a
calculation which predicts a non-linear leading correction, an $a\log a$ in
this particular case.

\subsubsection{Self-consistent treatment: linearity\label{scl}}

We now come back to the full equation (\ref{8-3}) and discuss the properties
of the self-consistent solution $W(\mathbf{k})$ . We define the dimensionless
function $v(x)$ of the dimensionless variable $x=k\lambda^{2}/a$ by:
\begin{equation}
W(\mathbf{k})=\frac{a^{2}}{\lambda^{2}}~~v\left(  \frac{k\lambda^{2}}%
{a}\right)  \label{essai}%
\end{equation}
Inserting this into (\ref{8-3}) provides:
\begin{align}
v(x)  &  =\frac{x^{2}}{4\pi}-\frac{8}{(2\pi)^{6}}\int d^{3}x^{\prime}\,\int
d^{3}y\,\frac{(a/\lambda)^{2}}{e^{(a/\lambda)^{2}v(\mathbf{x}^{\prime})}%
-1}~\frac{(a/\lambda)^{2}}{e^{(a/\lambda)^{2}v(\mathbf{x}^{\prime}%
-\mathbf{y})}-1}\times\nonumber\\
&  \times\left[  \frac{(a/\lambda)^{2}}{e^{(a/\lambda)^{2}v(\mathbf{x}%
+\mathbf{y})}-1}-\frac{(a/\lambda)^{2}}{e^{(a/\lambda)^{2}v(\mathbf{y})}%
-1}\right]  \label{self-dim0}%
\end{align}
If we take the limit $a/\lambda\rightarrow0$ of the integrand on the right
side of this equation (an operation justified below), we obtain the
parameter-free self-consistent equation:
\begin{equation}
v(x)=\frac{x^{2}}{4\pi}-\frac{8}{(2\pi)^{6}}\int d^{3}x^{\prime}\,\int
d^{3}y\,\frac{1}{v(\mathbf{x}^{\prime})v(\mathbf{x}^{\prime}-\mathbf{y}%
)}\left[  \frac{1}{v(\mathbf{x}+\mathbf{y})}-\frac{1}{v(\mathbf{y})}\right]
\label{self-dim}%
\end{equation}
(one can easily check that no other choice for the powers of $a$ and $\lambda$
in the scaling factors of (\ref{essai}) would allow the same complete
disappearance of all parameters from the equation).

We now assume\footnote{This assumption is supported by numerical and
analytical calculations of ~\cite{BBHLV-2}.} the existence of a solution to
(\ref{self-dim}) that, for values of $x\gg1$, is dominated by the free
spectrum term $x^{2}/4\pi$.\ We note that the integral is made ultraviolet
(UV) convergent by the difference that appears in the brackets.\ As for
infrared (IR) convergence, for small $x$'s the self-consistent solution will
automatically have a harder variation than $x^{2}$ (free spectrum) in order to
make the integral convergent, a variation in $x^{3/2}$ for instance\footnote{A
simple power counting argument applied to equation (\ref{self-dim}) predicts a
$x^{3/2}$ dependence of $v(x)$ for small values of $x$.} \cite{PP}.\ Being
convergent, the value of the integral can be obtained with arbitrary accuracy
from a finite domain of the dimensionless variable $x$, corresponding to
$\left\vert x\right\vert <D$ (where $D$ is a pure number).\ Coming back to
(\ref{self-dim0}), we see that the exponent in the denominator is smaller or
equal to $(a/\lambda)^{2}v(D)$, which tends to zero when $a/\lambda
\rightarrow0$ (since $v(D)$ is a pure number, independent of $a$).\ This, in
retrospect, justifies the lowest order expansion of the exponentials inside
the integrand which led us to (\ref{self-dim}), at least for values of the
current variable $x$ which are not too large (see below).

Finally, inserting (\ref{essai}) into (\ref{nc}) provides:%
\begin{equation}
\Delta n_{c}=-\frac{2}{\pi\lambda^{2}}\int dk~\frac{(a/\lambda)^{2}%
v(k\lambda^{2}/a)-(k\lambda)^{2}/4\pi}{(a/\lambda)^{2}v(k\lambda^{2}/a)}
\label{delta-nc}%
\end{equation}
or, with a variable change $x=k\lambda^{2}/a$:%
\begin{equation}
\Delta n_{c}=-\frac{2a}{\pi\lambda^{4}}\int dx~\frac{v(x)-x^{2}/4\pi}{v(x)}
\label{changt-var}%
\end{equation}
This result is equivalent to (\ref{nc}) and shows that the density shift at
constant temperature is indeed linear in $a$; consequently, the same property
is true for the temperature shift at constant density.

The values of the dimensionless variable $x$ that contribute to the integral
are smaller or comparable to $D$.\ In terms of the initial momentum variables
appearing in (\ref{8-3}), this domain has a range given by some number
multiplying $k_{c}$ defined in (\ref{9}). Going from (\ref{self-dim0}) to
(\ref{self-dim}) actually also requires that the current variable $x$ not be
too large, namely that $v(x)\ll(\lambda/a)^{2}$. Since, for large $x$, $v(x)$
is dominated by the kinetic energy term $\sim x^{2}$, this condition
corresponds to $x\ll\lambda/a$ and therefore to $k\lambda\ll1$. In other
words, the values of $\Sigma_{\mathbf{k}}$ obtained from (\ref{self-dim}) are
correct, but only for momenta significantly smaller that the maximum thermal
momentum. This caveat is irrelevant for the calculation of the leading term of
the density shift.\ As we will see in more detail in \S \ \ref{spectrum}, what
is important is the crossover value of $k$ at which the function $v(x)$
switches from a behavior dominated by the integral to a behavior dominated by
the quadratic kinetic energy.\ From (\ref{self-dim}), we know that this
phenomenon takes place for some finite value of $x\,$ which corresponds, in
terms of $k$, to a multiple of $k_{c}$. Finally, it is easy to see in
(\ref{9}) that, for sufficiently small values of $a$, the product
$k_{c}\lambda$ remains much smaller than $1$, which means that only particles
with low momentum have a role in the change of critical density.

The conclusion is that the non-linear self-consistent equation for the
effective energies leads to linear $a$ dependences of the temperature shift,
but only if it is treated self-consistently; otherwise, more complicated
spurious variations are introduced - see also ref. \cite{HR} for other
numerical illustrations of this property.

\subsection{Generalization}

We now go beyond the simple approximation (\ref{7}) for the effective energy
and generalize the considerations of the preceding section. In the complete
theory \cite{BBHLV-2, BBHLV-1}, the right side of this equation contains an
infinite sum of integrals, corresponding to all diagrams in the theory.\ They
start from the second order term in $a/\lambda$ already contained in
(\ref{7}), and continue with terms of all orders in $a/\lambda$, containing
higher dimensional integrals with more elaborate structure. The problem then
becomes significantly more complicated; we will see, nevertheless, that most
of the conclusions of the preceding section remain valid.\ We first discuss
the general relation between infrared convergence of the theory and the
linearity in $a$ of the leading term in the correction, then show that the
linearity of the simple model remains valid within the general theory, and
finally emphasize the difficulties arising in the precise calculation of the
linear coefficient $c$.

\subsubsection{Linearity and infrared convergence\label{IR}}

It is well known that the perturbative treatment of second order phase
transitions gives rise to infrared (IR)\ divergences. Here, we discuss the
close relation between these divergences and a possible breakdown of the
linearity in $a$ of the critical temperature shift.

Let us first come back to the single iteration approximation discussed in
\S \ \ref{spurious}.\ If we introduce again the effective chemical potential
$\mu_{\text{eff}}=\mu-4a\lambda^{2}n/\beta$, \ we see that this approximation
involves integrals - right side of (\ref{7}) - which, because they contain the
ideal gas spectrum $e_{\mathbf{k}}\sim k^{2}$, are IR\ divergent if
$\mu_{\text{eff}}=0$. More precisely, when $\mu_{\text{eff}}\rightarrow0$, one
obtains $\beta\Sigma_{0}-4a\lambda^{2}n\sim(a/\lambda)^{2}\log(-\beta
\mu_{\text{eff}})$, so that the critical condition (\ref{8-2}) leads to
logarithms in the value of the critical chemical potential, $-\beta
\mu_{\text{eff}}^{c}\sim(a/\lambda)^{2}\log(\lambda/a)$.\ But, for small $k$,
the integrand can be written as:%
\begin{equation}
\left[  \beta(e_{\mathbf{k}}-\mu_{\text{eff}})\right]  ^{-1}=\beta^{-1}\left[
\left(  \hbar^{2}k^{2}/2m\right)  -\mu_{\text{eff}}\right]  ^{-1}\sim\left[
\left(  k\zeta\right)  ^{2}+1\right]  ^{-1} \label{fraction}%
\end{equation}
where the mean field correlation length $\zeta$ is defined by $-\mu
_{\text{eff}}=\hbar^{2}/2m\zeta^{2}$, as in refs. \cite{BBHLV-2, BBHLV-1}.\ We
then see that $\zeta$ plays the role of a scaling factor for the $k$
dependence of the energies:
\begin{equation}
W(\mathbf{k})\sim\frac{a^{2}}{\lambda^{2}}v(k\zeta), \label{equation}%
\end{equation}
where $v(x)$ is a dimensionless function. We can finally use this result to
calculate the critical density shift, eq.~(\ref{nc}), and obtain $\Delta
n_{c}\propto\zeta^{-1}$, which is proportional to $(a/\lambda^{2})\sqrt
{\log(\lambda/a)}$.

For extremely small values of $a$, the logarithmic dependence on $a$ of the
correlation length then dominates the shift of the critical temperature.\ This
explains why the table of \cite{BBHLV-2} gives values of $c$ that depend
slightly on the ratio $a/\lambda$. Nevertheless, this logarithmic scaling is
spurious, an artifact of the truncation of the self-consistent equations at
the first iteration; as we have seen in \S \ \ref{scl}, the complete
self-consistent solution leads to a shift of $T_{c}$ linear in $a$.

More generally, the relation between IR convergence and linearity is easy to
understand.\ The critical condition (\ref{8-2}) is the key to the calculation
of the scaling factor $\zeta_{c}$ at the critical point, which in turn
determines the critical density. Since $\Sigma_{0}$ can be written as:%
\begin{equation}
\Sigma_{0}=\frac{4a\lambda^{2}n}{\beta}-\left(  \frac{a}{\lambda}\right)
^{2}~I(\zeta) \label{i-zeta}%
\end{equation}
where $I(\zeta)$ is a positive integral depending on the chemical potential
through $\zeta$, the critical condition (\ref{8-2}) reads:
\begin{equation}
-\mu_{\text{eff}}=\frac{\lambda^{2}}{4\pi\beta\zeta^{2}}=\left(  \frac
{a}{\lambda}\right)  ^{2}~I(\zeta) \label{mu-eff}%
\end{equation}
Graphically, as shown in figure 4, the critical value of $\zeta^{-2}$ is
obtained by intersecting a straight line with large slope $(\lambda/a)^{2}$
with the function $I\,$, considered as a function of $\zeta^{-2}$.\ The
intersection point is close to the vertical axis.\ If $I$ has a finite limit
$I_{0}$ when $\zeta^{-2}\rightarrow0$, we immediately obtain $\zeta^{-2}%
\sim\beta I_{0}~a^{2}/\lambda^{4}$, and from this simple scaling factor
linearity in $a$ follows.\ On the other hand, if this function diverges when
$\zeta^{-2}\rightarrow0$, the scaling factor has a more complicated $a$ dependence.

\begin{figure}[ptb]
\begin{center}
\includegraphics[
height=5cm, width=5.5cm ]{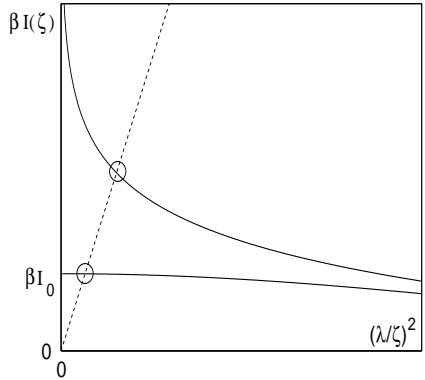}
\end{center}
\caption{Graphical construction to obtain the mean field correlation length,
$\zeta$; the lowest full line corresponds to the absence of IR divergences,
the upper full line to an IR diverging function. The slope of the dashed line
is proportional to $(\lambda/a)^{2}$ and diverges in the limit $a\rightarrow
0$.}%
\label{fig-6}%
\end{figure}

Ref. \cite{BBHLV-2} discusses various models illustrating the relation between
the absence of IR\ divergences and the linearity in $a$ of the leading term in
the correction. For instance, the \textquotedblleft bubble
sum\textquotedblright, or the \textquotedblleft ladder sum\textquotedblright,
both contain integrands with fractions that automatically control the
divergences, and therefore lead to linear dependences of the density shift in
$a$. This property explains why ref.\ \cite{Stoof-1} predicts a linear $a$
dependence, in an approach based on the many body $T$ matrix approximation,
similar to a ladder sum approximation.\ Clearly, these results do not
establish the linearity in $a$ for the full theory, since there is no special
reason why only bubbles or ladder should be retained from the full
series.\ The disappearance of the IR divergences from the bubble or ladder
sums is a special property of these subseries.\ For the reasons discussed
above, one could expect the density corrections to be dominated by other
diagrams containing IR\ divergences, therefore providing a result in $a\ln a$
for instance (this is indeed the result obtained later in ref. \cite{Stoof-2}%
), and making the contribution of isolated ladders negligible. We now discuss
the full theory to examine why the linearity in $a$ is actually preserved.

\subsubsection{Linearity in the full theory}

The reasoning of \S \ \ref{scl} can be extended to the more complicated
structure of the effective energies in the complete theory.\ We refer the
reader to the discussion given in \S \ 4 of ref.\ \cite{BBHLV-2}; here, we
give only a brief discussion of the argument leading to linearity.\ The
various terms in $\Sigma_{\mathbf{k}}$, corresponding to a series of diagrams
with increasing powers of $a/\lambda$, may be obtained by recurrence from one
order to the next. At each step, one includes one more power of $a/\lambda$,
one more $\mathbf{k}$ integral with a factor $\lambda^{3}$, and two more
$\rho_{\mathbf{k}}$'s, introducing two fractions with exponentials of
$W(\mathbf{k})$ in the denominator.\ Now, if we make the same change of
variable $x=k\lambda^{2}/a$ as before, we see that the change of integration
variable introduces an additional factor $(a/\lambda^{2})^{3}$, which comes in
addition to an $a/\lambda^{2}$ from the two coefficients above; altogether, we
get precisely the factor $(a/\lambda^{2})^{4}$ that is necessary to add a
factor $(a/\lambda^{2})^{2}$ to the numerator of each fraction, exactly as in
(\ref{self-dim0}).\ As a consequence, we obtain again a parameter free
non-linear equation, generalizing (\ref{self-dim}).

This new integral is more ultraviolet convergent than in the right side of
(\ref{self-dim}) since, for large $x$, two additional factors of $x^{2}$
overcompensate the additional integration factor $d^{3}x$.\ As for IR
convergence, again it is obtained because the solution $v(x)$ has to adapt its
small $x$ behavior to ensure it.\ Finally, the reasoning of \S \ \ref{scl} can
be made again with the more complicated non-linear equation, so that linearity
is obtained in this case also\footnote{In our proof, we have taken for granted
the existence of a self-consistent expression for the effective energies, as
in Green's function theory. The other assumption is the existence of a
solution to the non-linear equation.}.

This result is another illustration of the general property mentioned in the
preceding section: here, it is the self-consistent character of the
calculation that avoids the IR\ divergences - the solution automatically
adjusts its low $k$ behavior - and linearity is indeed obtained.

\subsubsection{Calculation of the linear coefficient $c$}

Even when linearity is proved, a difficult problem remains: the calculation of
the linear coefficient $c$.\ Analytically, this is an intricate problem,
because of its highly non-perturbative character. In the context of non-linear
self-consistent equations, we have seen that the problem is non-perturbative
at two levels: the choice of the non-linear equation, and the resolution of
the equation by successive approximations.\ In usual perturbation theory,
infrared divergences occur in all the integrals contained in the expression of
$\Sigma_{\mathbf{k}}$, preventing any expansion in the small parameter
$a/\lambda$ around the ideal gas spectrum $W(\mathbf{k})=e_{\mathbf{k}}%
$.\ When one includes higher and higher order in $a/\lambda$ terms in the
energy, these IR\ divergences become more and more severe, so that higher
powers of $a/\lambda$ in the denominator compensate exactly those in the
numerator.\ As a result, all terms are comparable - see the more detailed
discussion of refs. \cite{BBHLV-2, BBHLV-1}, which show that diagrams of all
orders in $a/\lambda$ may contribute comparable amounts to the small momentum
part of the spectrum, and therefore to the critical temperature shift. The
complexity of the general problem is illustrated by the second leading
correction: instead of being proportional to $a^{2}$, as one could naively
expect, it is proportional to $a^{2}\log a$, and therefore manifestly
non-analytic \cite{BBHL, AMT}.\ 

However, if we are only interested in the leading order (linear) shift of
$T_{c}$, a simplification is that factors $~\left[  e^{\beta(e_{\mathbf{k}%
}-\mu+\Sigma_{k})}-1\right]  ^{-1}$ can systematically be replaced by $\left[
\beta(e_{\mathbf{k}}-\mu+\Sigma_{k})\right]  ^{-1}$; in other words, the
problem reduces to classical field theory, a case in which the temperature
shift is exactly proportional to $a$ to all orders \cite{BBHLV-1}. This
simplification can be regarded as the generalization to all higher order
contributions of the approximation of (\ref{self-dim0}) by (\ref{self-dim}%
).\ In the absence of non-linear $a$ terms, numerical calculations for finite
values of $a$ are in principle easier, since a sometimes delicate
extrapolation to zero $a$ values becomes unnecessary.\ Nevertheless, in
practice, space discretization introduces another length, the lattice
parameter, which may re-introduce non-linearities; a careful extrapolation of
the lattice parameter to zero is then necessary.\ Taking this into account,
numerical lattice calculations \cite{AM,KPS} have indeed provided a precise
value, $c\simeq1.3$, which appears to be the best determination to date of
this coefficient. On the other hand, classical field theory does not offer an
analytic solution, since it is a priori not possible to select a class of well
defined perturbation diagrams that are sufficient to give a reasonable value
for $c$.

Finally, another approach to the problem is given by the \textquotedblleft
large $N$\textquotedblright\ expansion, where one studies the critical
temperature of a gas containing particles with many internal states, resulting
in an order parameter with $N$ components; the problem is exactly soluble in
the limit $N\rightarrow\infty$ \cite{GBZ} and the calculation of various
$1/N^{p}$ corrections \cite{GBZ, AT-2} allows one to extrapolate to the case
where the particles have only one internal state ($N=2$).\ Other
perturbative-variational methods have also been used, see
\cite{Souza-1,Souza-2,BR, Kleinert,Kastening-1,Kastening-2}.

\subsection{Discussion}

We have already mentioned in \S \ \ref{non-linear} that a spatial
rearrangement of atoms with low energies takes place, a similar effect to that
discussed by Lamb et al. \cite{Lamb}, but in the space of the relative
positions of the particles (correlations) instead of their ordinary
positions.\ The effect is maximal at $\mathbf{\ k}=0$, so that the atoms in
this level condense more easily than they would in a simple repulsive mean
field with no $\mathbf{k}$ dependence; of course, atoms with very low
$\mathbf{k}$ undergo a similar effect, which vanishes progressively when their
momentum increases beyond $k_{c}$.\ The resulting spectrum $e_{\mathbf{k}%
}+\Sigma_{\mathbf{k}}$ is shown schematically in figure 5-a, with the origin
of the energies at $\Sigma_{0}$ (the value of the chemical potential at
transition); for comparison, a purely parabolic spectrum is also shown (dashed
line).\ Figure 5-b shows the corresponding variations of the populations
$\rho_{\mathbf{k}}$, multiplied by $k^{2}$ in order to give directly their
contribution to the total population $N$.\ At condensation, when
$\mathbf{k}\rightarrow0$, the function goes to a constant for the ideal gas
(dashed line), but to zero for the ideal gas (full line) - in the presence of
interactions, the delta function corresponding to the presence of a
Bose-Einstein condensate appears with no pedestal.\ The comparison between the
situations for the ideal gas and the repulsive gas immediately shows that
every $\mathbf{k}\neq0$ level is less populated at the critical point in the
presence of interactions.\ Therefore, $\Delta n_{c}$ is negative at constant
$T$, or conversely $\Delta T_{c}$ is positive at constant $n$%
\footnote{Mathematically, a negative value is not totally excluded, but it
would require complicated compensation effects and somewhat unplausible $k$
variations of $\Sigma_{\mathbf{k}}$.}.

\begin{figure}[ptb]
\begin{center}
\includegraphics[
height=6cm, width=12cm ] {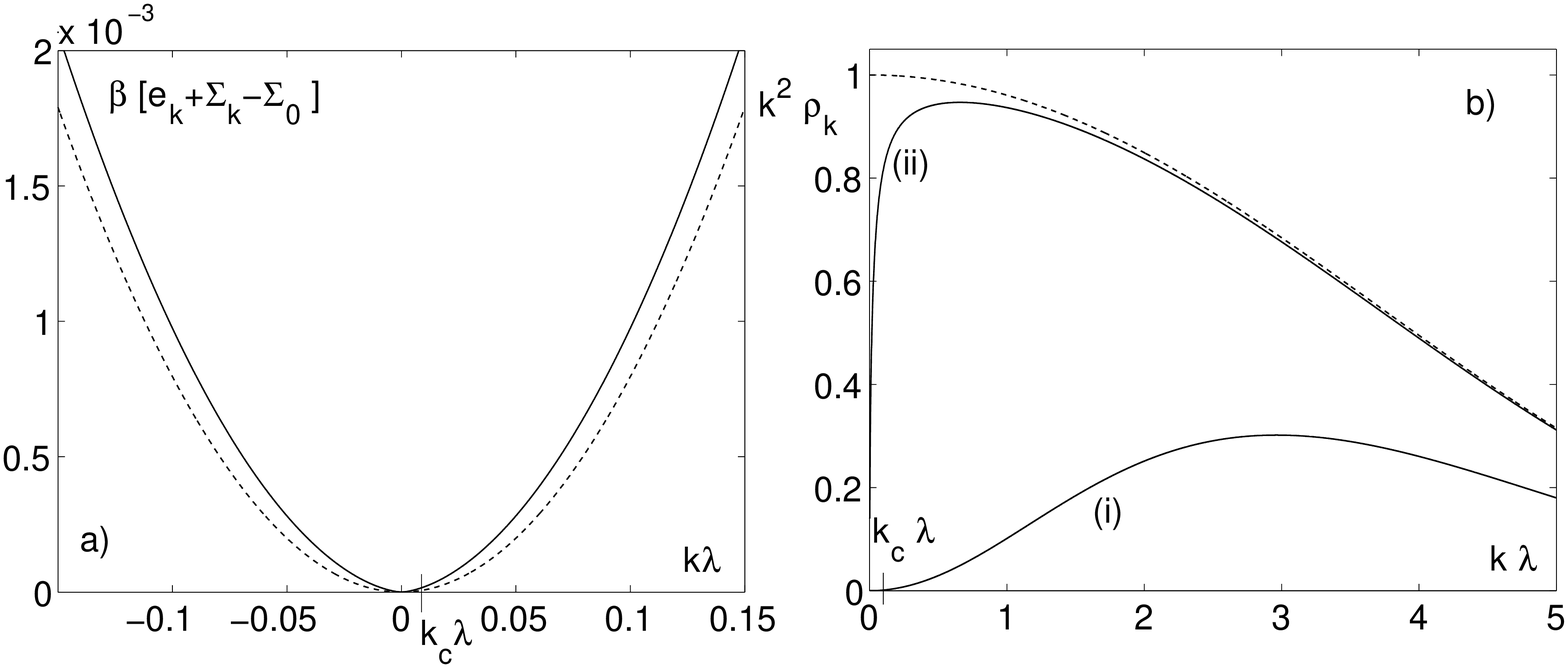}
\end{center}
\caption{Figure (a): plot of the function $e_{\mathbf{k}}+\Sigma_{\mathbf{k}}$
(full line), with the origin of the energies at $\ \Sigma_{0}$, and of the
same function for the ideal gas (dashed line). Figure (b) shows the
populations $\rho_{\mathbf{k}}$, multiplied by $k^{2}$, far above Bose
Einstein condensation (curve (i)), and exactly at the condensation point
(curve (ii)); again, the ideal gas curve is shown by a dashed line. Because of
the modified spectrum, the curve goes to zero at the origin in the presence of
repulsive interactions, unlike the ideal gas.}%
\label{fig-3}%
\end{figure}

\subsubsection{Role of the spectrum\label{spectrum}}

As in \cite{BBHLV-2, BBHLV-1}, we introduce the function $U(\mathbf{k})$ by:%
\begin{equation}
U(\mathbf{k})=\frac{2m}{\hbar^{2}}\left[  \Sigma_{\mathbf{k}}-\Sigma
_{0}\right]  =(2\pi\lambda^{2})^{-1}\left[  W(\mathbf{k})-\beta e_{\mathbf{k}%
}\right]  \label{11}%
\end{equation}
If we replace in (\ref{6}) $\rho_{\mathbf{k}}$ by the long wavelength
approximate form $\left[  \beta\left(  e_{\mathbf{k}}+\Sigma_{\mathbf{k}%
}-\Sigma_{0}\right)  \right]  ^{-1}$, we can express the change of the
critical density created by the interactions as:%
\begin{equation}
\Delta n_{c}\simeq-\frac{2}{\pi\lambda^{2}}\int dk~\frac{U(k)}{k^{2}+U(k)}
\label{12}%
\end{equation}
A priori, one could imagine that the value of $\Delta n_{c}$ depends crucially
on the details of the spectrum near $\mathbf{k}=0$. Actually, the density
change depends mostly on one parameter, the crossover point where $U(k)$
becomes equal to $k^{2}$ (the point at which the energy shift introduced by
the interaction equals the kinetic energy); it is relatively insensitive to
the details of the variations of the spectrum.\ For very low values of $k$
($\ll k_{c})$, the role of the function $U(k)$ is to harden the free particle
spectrum so that integrals such as (\ref{8-3}) converge; therefore $U(k)\gg
k^{2}$ and the integrand in (\ref{12}) is almost unity.\ For a value $k_{c}$
of $k$ comparable to $a/\lambda^{2}$, see equation (\ref{9}), there is a
crossover between the two functions, and one can expect that one rapidly
reaches the opposite regime where $k^{2}\gg U(k)$ and the integrand in
(\ref{12}) is almost zero.\ Altogether, we almost find an \textquotedblleft
all or nothing\textquotedblright\ regime where the population depletion jumps
rapidly from complete to zero, so that:%
\begin{equation}
\Delta n_{c}\simeq-\frac{2}{\pi\lambda^{2}}k_{c} \label{13}%
\end{equation}
which depends only on the value of $k$ at the crossover and not on the details
of the variations of $U(k)$. This explains in particular why the calculation
of the critical density, or of the critical temperature, is relatively
insensitive to the exact value of the universal coefficient $\eta$ which
characterizes the power dependence of $U(k)\sim k^{2-\eta}$ at low $k$.

\subsubsection{Role of exchange cycles}

Feynman has emphasized the role of exchange cycles in Bose-Einstein
condensation \cite{Feynman-1, Feynman-2}: condensation takes place when a
macroscopic exchange cycle can propagate across the whole sample - see also
Elser's work \cite{loop-gas}, as well as \cite{Ceperley} for a review.\ The
notion of exchange cycles arises in first quantization, where identical
particles are numbered, and where a symmetrization operator $S$ is applied in
a second step to obtain the correct physical states.\ This operator can be
decomposed into cycles of particles, usually visualized as a closed path
joining all the particles involved \cite{Ceperley}. The probability of
occurrence of a given path decreases rapidly when the distance between two
consecutive particles much exceeds the thermal wavelength $\lambda$; this
provides the length scale for the maximum jump of a cycle between particles

Reference \cite{GCL} discusses how the shift of the critical temperature
introduced by the interactions in a homogeneous system can be interpreted in
terms of exchange cycles.\ In an ideal Bose gas just above the transition
point, the density fluctuations are large because exchange effects tend to
bunch the particles together.\ There will be regions of space where the
density is lower than average and where, because of the maximum length
$\lambda$ of the jump of an exchange path, cycles will not easily propagate;
clearly, this effect will oppose the appearance of Bose-Einstein
condensation.\ On the other hand, in a dilute repulsive gas, mutual repulsions
create a restoring force which tends to make the system more homogeneous; the
occurrence of large regions of low density fluctuations becomes unlikely. This
will facilitate the propagation of large exchange cycles across the whole
system, i.e., Bose-Einstein condensation. Therefore, a lower density (or
conversely a higher temperature) are needed to reach
condensation\footnote{This is actually an interesting situation: in most
cases, repulsive interactions between particles tend to mask quantum effects,
especially exchange effects (for instance, hard core potential effects reduce
exchange effects in liquid helium three and four); for the dilute system here,
they actually enhance them.\ }.\ 

At this point, one may wonder what is the relation between this physical
explanation and the calculations presented above: what is relevant is the
effects of IR divergences, and therefore physical properties of the system at
very large distances; in this section, what matters is the correlations of
particles at a distance of the order of $\lambda$.\ This apparent paradox is
removed if one remembers that, in our preceding discussion, we were studying
one-body properties; here we are dealing with the propagation of exchange
cycles, which depends on two-body properties.\ A phenomenon which appears at
long distances for single particles can translate into shorter range
properties for correlations. Here, we will not give a precise discussion of
this point, but just develop a plausibility argument.\ We use a simplified
model where the two-body density operator $\rho_{II}$ in the gas is given by:%

\begin{equation}
\rho_{II}(1,2)=\left[  \rho_{I}(1)\otimes\rho_{I}(2)\right]  \left[
1+P_{\text{ex}}\right]  +.... \label{14}%
\end{equation}
where $\rho_{I\text{ }}$ is the one-body density operator and $P_{\text{ex}}$
is the exchange operator between particles $1$ and $2$.\ This relation is
exact for the ideal gas, but only an approximation for the interacting
gas.\ Even if some effects of the interactions are already contained in
(\ref{14}) through changes of the $\rho_{I}$'s, it is clear that more terms
should be added to the right of (\ref{14}) to get an exact formula - in
particular terms that introduce short range correlations between the
particles.\ This approximation is sufficient for our present qualitative discussion.

The direct term in the right side of (\ref{14}) is simply a product containing
no spatial correlation, so that we will ignore it and focus on the exchange
term containing $P_{\text{ex}}$ with diagonal elements:
\begin{equation}
<\mathbf{r}_{1},\mathbf{r}_{2}\mid\rho_{II}^{\text{ex}}(1,2)\mid\mathbf{r}%
_{1},\mathbf{r}_{2}>\,=\,<\mathbf{r}_{1},\mathbf{r}_{2}\mid\rho_{I}%
(1)\otimes\rho_{I}(2)\mid\mathbf{r}_{2},\mathbf{r}_{1}> \label{15}%
\end{equation}
Since one-particle density operators are diagonal in the momentum
representation (translational invariance), this term is proportional to:
\begin{equation}
\int d^{3}k_{1}\int d^{3}k_{2}\,\rho_{\mathbf{k}_{1}}\rho_{\mathbf{k}_{2}%
}\,e^{i(\mathbf{k}_{1}-\mathbf{k}_{2})\cdot(\mathbf{r}_{1}-\mathbf{r}_{2})}
\label{16}%
\end{equation}
Therefore the two-body correlation function is independent of the sum
$\mathbf{r}_{1}+\mathbf{r}_{2}$ (position of the center of mass of the two
particles), while its $\mathbf{r}_{1}-\mathbf{r}_{2}=\mathbf{r}$ dependence is
given by the Fourier transform of the integral:
\begin{equation}
\int d^{3}K\,\,\,\,\,\rho_{(\mathbf{K}+\mathbf{k})/2}\,\times\,\rho
_{(\mathbf{K}-\mathbf{k})/2} \label{17}%
\end{equation}
which, using parity, can be expressed as a convolution integral:
\begin{equation}
\int d^{3}K\,\,\,\,\,\rho_{(\mathbf{k}+\mathbf{K})/2}\,\times\,\rho
_{(\mathbf{k}-\mathbf{K})/2} \label{18}%
\end{equation}

At the critical point, we have seen above that the distribution function
$\rho_{\mathbf{k}}$ is the sum of two components: a Bose-Einstein distribution
obtained within a mean-field approximation with a slightly negative value of
the effective chemical potential (proportional to the square of the scattering
length $a$), and a critical perturbation introduced for small $k$ (comparable
to $k_{c}$) by the correlation effects; the latter is difficult to calculate,
but small.\ We therefore have:
\begin{equation}
\rho_{\mathbf{k}}=\rho_{\mathbf{k}}^{(0)}+\Delta\rho_{\mathbf{k}} \label{rho6}%
\end{equation}
where the first term has a width comparable to $1/\lambda$ while the second
has a width $k_{c}$.\ The two-body correlation function now appears as the sum
of three terms: a term in $\left[  \rho_{\mathbf{k}}^{(0)}\right]  ^{2}$, a
crossed term in $\rho_{\mathbf{k}}^{(0)}\times\Delta\rho_{\mathbf{k}}$ on
which we will focus our interest, and finally a term in $\left[  \Delta
\rho_{\mathbf{k}}\right]  ^{2}$, which we will ignore since it is smaller.

The first term is well-known \cite{Van-Hove}: it corresponds essentially to an
ideal Bose gas close to the transition point and contains the usual
\textquotedblleft exchange bump\textquotedblright, with a width comparable to
$\lambda$, as well as a long exponential tail\footnote{The tail can easily be
obtained by using the approximation $\rho_{\mathbf{k}}^{(0)}\sim\beta
^{-1}(e_{\mathbf{k}}-\mu)^{-1}$, equivalent to a zero Matsubara frequency
approximation - see for instance exercise 12.9 of \cite{Huang}.\ Since the
effective chemical potential is proportional to $a^{2}$ at the transition
point, the range of this exponential tail is of order $\lambda^{2}/a$.}.\ The
second term is the crossed term; its Fourier transform appears as the
convolution of the function $\rho_{\mathbf{k/2}}^{(0)}$ by $\Delta
\rho_{\mathbf{k}/2}$.\ Since $\Delta\rho_{\mathbf{k}/2}$ is much narrower than
$\rho_{\mathbf{k/2}}^{(0)}$, the function $\Delta\rho_{\mathbf{k}/2}$ may be
approximated by a delta function at the origin, with a positive weight
$d$.\ We then obtain for the Fourier transform of this term:
\begin{equation}
d\times\rho_{\mathbf{k}/2} \label{rho7}%
\end{equation}
This corresponds to a change of the spatial correlation function that is
positive; it has basically the same spatial dependence as the zero order term,
except that no squared function appears here.\ It therefore contains an
exchange bump that extends to distances comparable to the thermal wavelength,
but slightly further because it no longer includes a convolution of two $\rho
$'s; there still is a long exponential tail, with twice the range of the zero
order tail.

This simple model illustrates how the physics behind the reduction of the
critical density may appear at a different scale, depending whether it is
expressed in terms of single particle properties or in terms of
correlations.\ In the former case, the critical phenomenon is contained in
$\Delta\rho_{\mathbf{k}}$, which contains mostly small $\mathbf{k}$'s and is
dominated by long distances; in the latter, $\Delta\rho_{\mathbf{k}}$
disappears from the leading term, so that distances of the order of $\lambda$
remain relevant.

\section{Mean field and correlation effects combined in a trap\label{combined}%
}

As already mentioned, in the thermodynamic limit, Bose-Einstein condensation
is reached in a trap when the density at the center of the trap is exactly the
critical density for a uniform gas, including of course the corrections
introduced by the correlations.\ Two effects (the compressibility effect of
\S \ \ref{compressibility} and the correlation effect of
\S \ \ref{correlation}) then add to each other\footnote{More precisely,
subtract from each other, since they have opposite signs.} to determine the
change of the critical value of $N$\ at a given temperature (or, conversely,
the critical temperature at fixed $N$).\ Nevertheless, as remarked by Arnold
and Tomasik \cite{AT-1}, their combination is not trivial because spatial
effects in a trap mix up various orders in $a$.\ In this section we study the
effects of the interactions on the density profile of a gas in a trap,
assuming that the temperature is above, or just at the critical
temperature.\ We start from the number density at the center, from which we
can calculate the density profile by using an equation of state obtained
within mean field approximation.\ The critical density profile is then derived
by setting the density at the center to its critical value, including the
correlation effects discussed in the preceding sections, which are beyond mean
field theory.\ There is nevertheless no inconsistency in this approach; the
reason is that the calculation of the equation of state of the gas reduces to
an ordinary virial correction to the pressure, which is perturbative, while
the calculation of the critical density remains essentially non perturbative.

We will see that the specific properties of the curves of figure 1 play a role
in this problem, in particular the large compressibility of the gas at the
critical point.\ Roughly speaking, the compressibility of the gas at the
center of the trap is large ($\sim1/a$), but remains much smaller in most
other regions of the trap (practically independent of $a$).\ One can then
anticipate that a change of the pressure that is only second order in $a$ will
correspond to a change of the density of the same order in those regions, but
first order at the center.\ This will allow one to meet the change of the
condensation condition introduced by the correlations effects.\ But, since $N$
is primarily determined by the density of the gas in those low compressibility
regions, this change corresponds to a second order
variation\footnote{Conversely, one can assume that $N$ is fixed and that the
temperature is reduced from above the transition; the size of the atomic cloud
then decreases progressively.\ When the system reaches a temperature close to
condensation, the compressibility at the center becomes $\sim1/a$ and
therefore very large, so that particles tend to accumulate more around this
point of space than elsewhere in the trap.\ Consequently, a second order in
$a$ change of the temperature is sufficient to create a first order change of
the local density.} of $N$. We now discuss this question more precisely.

\subsection{Density profile}

In a first step, we consider the number density of the gas at the center of
the trap $n(0)$ as a free parameter.\ We come back to equation (\ref{4}) and
distinguish two cases:

(i) Classical gas ($n\lambda^{3}\ll1$)

The relation between the chemical potential and the density is then $\beta
\mu_{\text{eff}}\simeq\ln(n\lambda^{3})\,$, which implies $J^{^{\prime}}%
\simeq1/n\lambda^{3}$, so that (\ref{4}) becomes:%
\begin{equation}
-\nabla\left[  \beta V+4a\lambda^{2}n\right]  =\nabla\ln(n) \label{m1}%
\end{equation}
An $\mathbf{r}$ integration then gives:%
\begin{equation}
n\sim\exp\left[  -\beta V-4a\lambda^{2}n\right]  \label{m2}%
\end{equation}
which is nothing but the usual Boltzmann exponential. Around the center of the
trap, the density varies proportionally to the potential energy $V$:%
\begin{equation}
n(\mathbf{r})\simeq n(0)\left[  1-\frac{\beta V(\mathbf{r})}{1+4a\lambda
^{2}n(0)}+..\right]  \text{ \ (if }r\rightarrow0\text{)} \label{m2-bis}%
\end{equation}

(ii) Quantum non condensed gas ($1\lesssim n\lambda^{3}\leq\zeta_{3/2}$)

For simplicity, we assume that the gas is strongly degenerate at the center of
the trap ($n(0)\lambda^{3}$ is close to $\zeta_{3/2}$), and we limit our study
to the region of the trap where this degeneracy remains strong.\ We can then
use the Mellin formula limited to lowest order (a more precise calculation is
given in the appendix): $g_{3/2}(\exp\beta\mu)\simeq\zeta_{3/2}-2\sqrt
{-\pi\beta\mu}$ \cite{Pathria, Robinson}, where $\zeta_{3/2}$ is the value at
the origin of the $g_{3/2}$ function ($\zeta_{3/2}=g_{3/2}(z=1)\simeq
2.61..$).\ This provides:%
\begin{equation}
J\simeq-\frac{1}{4\pi}\left(  n\lambda^{3}-n_{c}^{0}\lambda^{3}\right)
^{2}+.. \label{m3}%
\end{equation}
\ where:%
\begin{equation}
n_{c}^{0}=\zeta_{3/2}~\lambda^{-3} \label{m5-bis}%
\end{equation}
is the critical density of the ideal gas. This result, inserted into
(\ref{4}), allows us to integrate the gradients on each side of the equation
and yields (assuming that the potential at the center of the trap vanishes):%
\begin{equation}
-\beta V=-\frac{\lambda^{6}}{4\pi}\left[  \left(  n-n_{c}^{0}\right)
^{2}-\left(  n(0)-n_{c}^{0}\right)  ^{2}\right]  +4\frac{a}{\lambda}\left(
n-n(0)\right)  \lambda^{3} \label{m4}%
\end{equation}
This equation is second degree in $n$; choosing the solution which tends to
$n(0)$ when $r\rightarrow0$ provides:%
\begin{equation}
n(\mathbf{r})=n_{c}^{0}+\frac{8\pi a}{\lambda^{4}}-\sqrt{\left[  n_{c}%
^{0}-n(0)+\frac{8\pi a}{\lambda^{4}}\right]  ^{2}+\frac{4\pi}{\lambda^{6}%
}\beta V(\mathbf{r})} \label{m5}%
\end{equation}
Equation (\ref{m5}) gives the density profile of the gas as a function of
$\mathbf{r}$, provided that the gas is in the quantum regime.

For an ideal gas ($a=0$), figure 6-a shows the density profiles obtained as a
function of the central density $n(0)$, assuming a quadratic potential
$V(\mathbf{r})\sim r^{2}$; as usual, the thermal range $R_{T}$ is
defined\footnote{$R_{T}$ is the size of the ideal classical gas in the trap at
temperature $T$.} by $(r/R_{T})^{2}=\beta V(r)$. If $n(0)=n_{c}^{0}$, the
density variations are linear in $r$ in all the quantum region; for lower
values of $n(0)$, there is a parabolic variation near the center of the trap,
in a domain that becomes larger and larger when $n_{c}^{0}-n(0)$ increases.

For an interacting gas ($a>0$), one can distinguish in (\ref{m5}) two
possibilities, depending whether the potential energy $\beta V$ is larger or
smaller than $\left[  \left(  n_{c}^{0}-n(0)\right)  \lambda^{3}+8\pi
a/\lambda\right]  ^{2}$.\ Near the center of the trap, $V$ is small and the
density variation is proportional to $V$ (quadratic in $\mathbf{r}$ for a
harmonic potential):%
\begin{equation}
n(\mathbf{r})\simeq n(0)-\frac{2\pi\beta V(\mathbf{r})}{\lambda^{3}\left[
(n_{c}^{0}-n(0))\lambda^{3}+(8\pi a/\lambda)\right]  }+..\text{ (if
}r\rightarrow0\text{)} \label{m6}%
\end{equation}
This formula is similar to (\ref{m2-bis}) but predicts significantly different
results.\ For instance, if $a$ is sufficiently small and if the central
density is sufficiently close to its critical value, the change of the density
induced by the potential becomes arbitrarily large - while it remains constant
for a classical gas; this is a consequence of the large compressibility of the
quantum gas at the origin. Further from the center of the trap, $V$ is large,
and (\ref{m5}) becomes:%
\begin{equation}
n(\mathbf{r})\simeq n_{c}^{0}+\frac{8\pi a}{\lambda^{4}}-\sqrt{\frac{4\pi\beta
V(\mathbf{r})}{\lambda^{6}}}\left[  1+\frac{\left[  (n_{c}^{0}-n(0))\lambda
^{3}+(8\pi a/\lambda)\right]  ^{2}}{8\pi\beta V(\mathbf{r})}+..\right]
\label{m7}%
\end{equation}
which now predicts a square root dependence in $V$. Figure 6-b illustrates
these two different behaviors of the density.

\begin{figure}[ptb]
\begin{center}
\includegraphics[
height=5cm, width=12cm ]{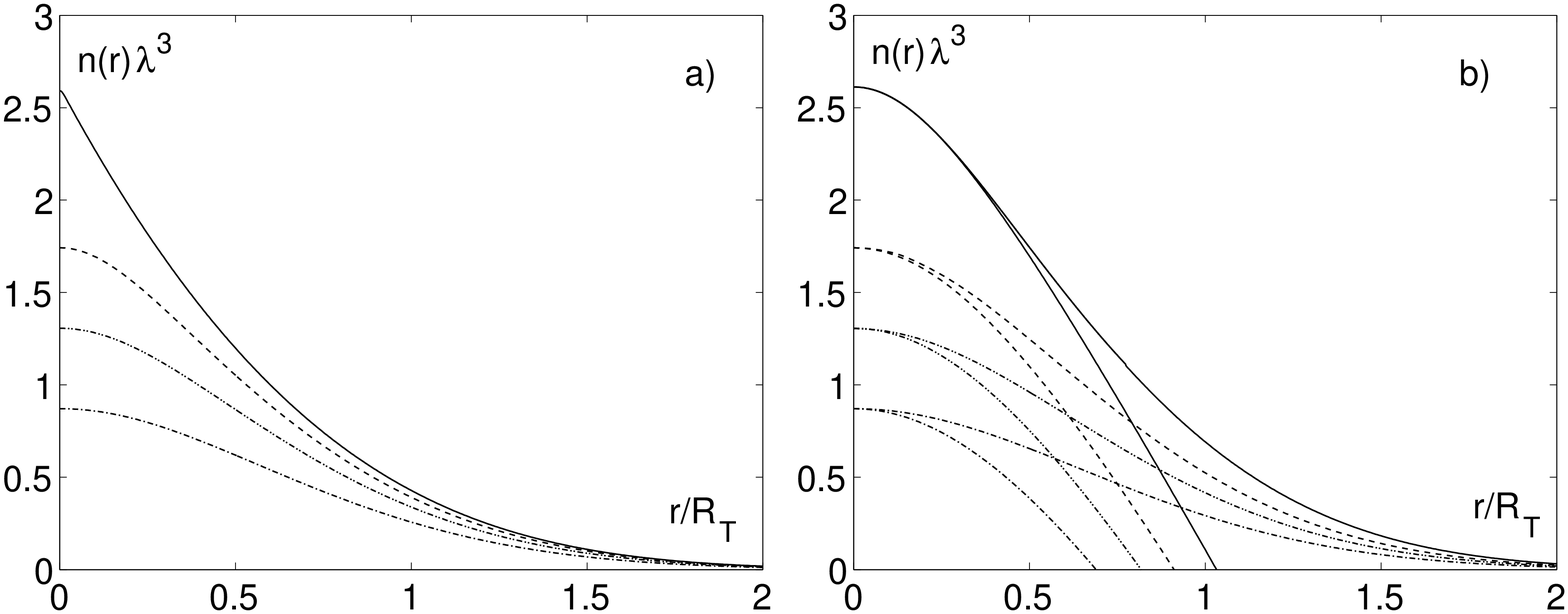}
\end{center}
\caption{ Figure (a) shows the modification of the density profile for the
trapped ideal gas when increasing the central density until it reaches the
condensation condition (full line: $n(0)\lambda^{3}=\zeta_{3/2}$; dashed line:
$n(0)\lambda^{3}=2\zeta_{3/2}/3$; dash-double dot line: $n(0)\lambda^{3}%
=\zeta_{3/2}/2$; dash-dot line: $n(0)\lambda^{3}=\zeta_{3/2}/3$). Figure (b)
shows the corresponding density profile for the trapped repulsive gas
calculated within mean field theory. The asymptotes at the center of the trap
are calculated with the lowest order approximation of the Mellin formula,
equation (\ref{m5}).}%
\label{fig-4}%
\end{figure}

\subsection{Critical density profile}

We are now in position to discuss the effects of the interactions on the
density profile at the transition point, at least in the central region of the
trap where the gas remains strongly degenerate.\ We set the central density to
its critical value $n_{c}$ and use the above equations to calculate
$n(\mathbf{r})$. This calculation is of course not exact, for different
reasons: first, (\ref{m5}) is only valid within mean field theory; second, the
use of the first correction only in the Mellin formula limits us to a
sufficiently degenerate gas.\ Both approximations could be improved: the first
by including second order perturbative corrections to the equation of state,
as done in ref.\ \cite{AT-1}; the second by including more terms in the Mellin
formula, as done in the appendix.\ Our calculation nevertheless remains
sufficient for a qualitative discussion.

Let us first ignore the effects of the correlations on the critical density at
the center.\ We then set $n(0)=n_{c}^{0}$ and see that the density corrections
are positive and first order in $a$ everywhere in the trap - second term in
the right side of eq. (\ref{m7}) - except in a small region around the center
where the density change is still positive but $\propto1/a$ - fraction in the
right side of eq. (\ref{m6}).\ For a harmonic potential, this small region has
a size proportional to $a$, so that the corresponding contribution to $N$ is
only second order in $a$. Therefore, the main contribution arises from all the
rest of the trap and provides a first order in $a$ change of $N$ (at constant
temperature); conversely, calculating the temperature shift at constant $N$
provides (\ref{5}), as shown in ref. \cite{GPS}.

We now take into account the effects of the correlations on the critical
density at the center of the trap.\ We then need to choose a slightly smaller
value of $n(0)$ since, according to (\ref{changt-var}), $n_{c}$ is reduced by
an amount that is first order in $a$. With respect to the calculation of the
preceding paragraph, eq. (\ref{m6}) predicts an additional negative change,
first order in $a$, of the density around the center of the trap.\ On the
other hand, eq. (\ref{m7}) predicts only a second order change induced by the
new value of $n(0)$; the crossover between the regions of space where these
equations apply is obtained when the potential energy is second order in $a$,
i.e., at a first order distance from the origin if the potential is harmonic.
This corresponds only to a fourth order correction to the total number of
particles $N$, because the volume of this region is proportional to $a^{3}$
for a harmonic potential; this correction remains therefore negligible when
compared to the second order in $a$ contribution from all other regions in the
trap.\ Altogether, we see that the additional change of $N$ introduced by the
correlation effects is therefore only second order in $a$; this is to be
compared with the first order correction introduced by mean field effects
contained in the term in $a/\pi\lambda^{4}$ on the right side of (\ref{m5}).

We therefore recover the conclusions of Arnold and Tomasik \cite{AT-1}: the
two $N$ changes subtract from each other, but not at the same order in
$a$.\ This offset of one order illustrates again the relatively loose
connection between the total number of atoms in the trap $N$ and the physics
at the center, where Bose-Einstein first takes place: most of the particles
contributing to $N$ are too far from condensation to play a role in it.\ If
the central density $n(0)$ were accessible, it would provide a more adequate
variable than $N$ for the study of the BEC\ phenomenon.\ See ref.\ \cite{HSC}
for a proposal of a method (adiabatic ramping down of the trap frequency)
aimed at enhancing the visibility of the correlation effects against the
background of mean field effects.

\section{Conclusion}

The calculation of the critical temperature of a dilute Bose gas has several
distinguishing features. One could of course try to go further than the
leading $a$ correction to the critical temperature and attempt to describe
analytically the full density variations obtained by path integral Monte Carlo
methods in \cite{GCL}.\ Quantitatively it is clear that at high densities the
blocking of exchange created by the repulsive hard core part of the potential
tends to reduce the critical temperature, an effect which goes in the opposite
direction to what happens for dilute gases. Ref. \cite{FW} gives a discussion
of this effect in terms of change of the effective mass of the particles,
arising from a $k$ dependence of the exchange term in the expression of the
mean field. It would be interesting to make this idea more quantitative and to
explore the specific details of BEC in all density regimes.

\bigskip

ACKNOWLEDGMENTS: We gratefully thank Peter Arnold and Masud Haque for several
useful discussions.\ Laboratoire. Kastler Brossel is \textquotedblleft
laboratoire associ\'{e} ENS-CNRS-UPMC UMR\ 8552\textquotedblright. Laboratoire
de Physique Th\'{e}orique des Liquides is \textquotedblleft laboratoire
associ\'{e} du CNRS et de l'UPMC Paris 6, UMR 7600\textquotedblright. One of
the authors (J.N.F.) is grateful to \textquotedblleft Association
Dephy\textquotedblright\ for a research grant and to \textquotedblleft
Soci\'{e}t\'{e} Fran\c{c}aise de Physique\textquotedblright for its support.
This reserach was supported in part by US National Science Foundation Grant PHY00-98353.

\begin{center}
APPENDIX
\end{center}

A better approximation for $g_{3/2\text{ }}$than that used in
\S \ \ref{combined} is:%
\begin{equation}
g_{3/2}(\exp\beta\mu)\simeq\zeta_{3/2}-2\sqrt{-\pi\beta\mu}-b~\beta\mu+...
\label{a-1}%
\end{equation}
with $b=1.46..$ \cite{Pathria, Robinson}. Adding a linear term to the square
root actually provides values which are reasonably accurate until $\beta\mu$
reaches values where the classical regime is obtained ($\beta\mu
<-1$).\ Inverting the relation $n\lambda^{3}=$ $g_{3/2}(\exp{\beta
\mu_{\text{eff}}})$ then gives:%
\begin{equation}
-\beta\mu_{\text{eff}}=\frac{\pi}{b^{2}}\left[  2+\frac{b}{\pi}\left(
n-n_{c}^{0}\right)  \lambda^{3}-2\sqrt{1+\frac{b}{\pi}\left(  n-n_{c}%
^{0}\right)  \lambda^{3}}\right]  \label{a-2}%
\end{equation}
which defines the function $J(n\lambda^{3})$. We therefore have:%
\begin{equation}
J^{^{\prime}}=-\frac{1}{b}\left[  1-\frac{1}{\sqrt{1+b\lambda^{3}\left(
n-n_{c}^{0}\right)  /\pi}}\right]  \label{a-3}%
\end{equation}
Relation (\ref{4}) then becomes, after integration of the gradients:%
\begin{equation}%
\begin{array}
[c]{ll}%
-\beta V & =4a\lambda^{2}\left(  n-n(0)\right) \\
& -\frac{\lambda^{3}}{b}\left[  \left(  n-n(0)\right)  -\frac{2\pi}%
{b\lambda^{3}}\sqrt{1+b\lambda^{3}\left(  n-n_{c}^{0}\right)  /\pi}+\frac
{2\pi}{b\lambda^{3}}\sqrt{1+b\lambda^{3}\left(  n(0)-n_{c}^{0}\right)  /\pi
}\right]
\end{array}
\label{a-4}%
\end{equation}
This equation provides a direct relation between the potential energy and the
local density.\ For instance, for a quadratic potential, the position $r$ is
obtained by taking the square root of the right side, multiplied by some
coefficient. This allows one to make a precise calculation of the density
profile in all regions of the trap where the gas is in a quantum regime.
Figure 7 gives a comparison between the density profile obtained in the mean
field approximation with (\ref{a-4}) and (\ref{m5}).

\begin{figure}[ptb]
\begin{center}
\includegraphics[
height=5cm, width=5.5cm ]{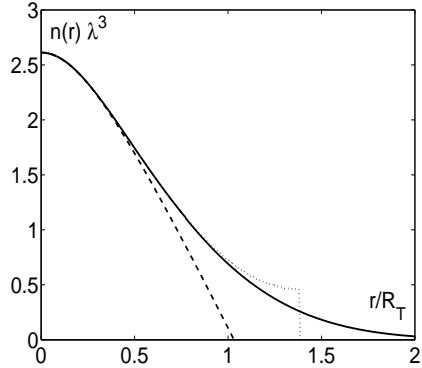}
\end{center}
\caption{Full line: mean field density profile calculated exactly (equation
(\ref{3-1})). Dashed line: same profile with the lowest order approximation of
the Mellin formula (equation (\ref{m5})). Dotted line: same profile with the
inclusion of the linear correction (equation (\ref{a-4})).}%
\label{fig-5}%
\end{figure}

\end{document}